# Heterogeneous Random Laser with Switching Activity Visualized by Replica Symmetry Breaking Maps

Loredana M. Massaro,[†] Silvia Gentilini,[‡] Alberto Portone,[¶] Andrea Camposeo,[¶] Dario Pisignano,[§] Claudio Conti,[‡] Neda Ghofraniha[*,‡]

[†]*Dipartimento di Fisica, Università La Sapienza, P.le A. Moro 5, I-00185, Rome, Italy*

[‡]*Istituto dei Sistemi Complessi-CNR, UOS Università La Sapienza, P. le A. Moro 2, I-00185, Rome, Italy*

[¶]*NEST, Istituto Nanoscienze-CNR and Scuola Normale Superiore, Piazza S. Silvestro 12, I-56127 Pisa, Italy*

[§]*NEST, Istituto Nanoscienze-CNR and Dipartimento di Fisica, Università di Pisa, I-56127 Pisa, Italy*

[*]E-mail: neda.ghofraniha@cnr.it

**ABSTRACT**. In the past decade, complex networks of light emitters are proposed as novel platforms for photonic circuits and lab-on-chip active devices. Lasing networks made by connected multiple gain components and graphs of nanoscale random lasers (RLs) obtained from complex meshes of polymeric nanofibers are successful prototypes. However, in the reported research, mainly collective emission from a whole network of resonators is investigated, and only in a few cases, the emission from single points showing, although homogeneous and broad, spatial emission. In all cases, simultaneous activation of the miniaturized lasers is observed. Here, differently, we realize heterogeneous random lasers made of ribbon-like and highly porous fibers with evident RL action from separated micrometric domains that alternatively switch on and off by tuning the pumping light intensity. We visualize this novel effect by building for the first time replica symmetry breaking (RSB) maps of the emitting fibers with 2 μm spatial resolution. In addition, we calculate the spatial correlations of the laser regions showing clearly an average extension of 50 μm. The observed blinking effect is due to mode interaction along light guiding fibers and opens new avenues in the fabrication of flexible photonic networks with specific and adaptable activity.

Wave guiding is known to be fundamental for the transmission of light and signals at the speed of light on long distances with low attenuation and a wide usable range of wavelengths. Besides kilometric transportation, photonic circuits are capable of integrating multiple functions on the same chip thanks to the guiding of light through interconnections between several different optical components.

In the last years, flexible polymeric nanofibers[1] have been fabricated and are proposed successfully for several photonic applications.[2-5] Indeed, networks of nanolasers have been realized from controlled meshes of dyed polymeric nanofibers.[6] Under pumping, the gain is amplified along the branches of the network and entrapped at the nodes due to multiple scattering, giving rise to efficient random lasers[7,8] and the opportunity to obtain low-cost, flexible, and scalable arrays of nano- and microlasers.[9] Similarly, a novel theoretical scheme named lasing network (LANER) has been proposed. It consists of a complex active optical network whose connectivity induces a form of topological disorder and can display laser action. The system is experimentally realized with optical fibers linked with couplers and with one or more coherently amplified sections.[10,11] In the reported research on complex networks of light emitters, the resonators are highlighted at the nodes or at points where additive nanoparticles are accumulated[12,13] and spectral emission from the whole network is detected[14,15] mostly from the edge of the planar devices, hindering the behavior of single emitters and their possible interactions.[13] In a few cases the RL emission from the different points of the system has been investigated, showing homogeneous and broad spatial emission.[12]

In this work, differently, we use ribbon-like and highly porous polymeric electrospun fibers to realize random lasers with heterogeneous stimulated emission both along the arms and at the nodes of the network. The emission does not come out homogeneously from the whole device, but micrometric domains lase separately and can alternatively switch on and off by changing the pumping intensity. To obtain this effect, the porous electrospun fibers are dyed, and once illuminated, the fluorescence is scattered in any part of the system and additionally guided along the arms. This joint effect leads to random laser emission in regions where the gain overcomes the losses, and the topology of these emitting domains changes at different levels of pumping due to mode competition along the active branches of the network. We show for the first time such an effect by direct imaging the RL intensity fluctuations and by building replica symmetry breaking (RSB) maps of the emitting fibers with a spatial resolution of 2 µm.

**Results and discussion**

**Heterogeneous intensity fluctuations.**

As an efficient heterogeneous branched random laser, a mesh of highly porous and ribbon-shaped fibers are realized by electrospinning dyed poly(methyl methacrylate) (PMMA) from solutions of chloroform and dimethyl sulfoxide (DMSO; volume ratio 6:1). Rhodamine-6G is then added to the solution with a 2.5% (wt/wt) with respect to the PMMA matrix, as final concentration. Electrospinning is performed by means of a 16 kV bias from a Glassman high voltage apparatus (series EL), applied between the needle of a 1 mL syringe containing the polymer solution and a $10 \times 10$ cm$^2$ metal plate. The needle-plate distance is as large as 22 cm, and the solution flow is kept at 4 mL h$^{-1}$. Scanning electron microscopy SEM inspection (Merlin, Zeiss) highlights fibers with a ribbon width of 15.0±1.5 µm, as reported in Figure 1a. The scattering capacity and light entrapment along the arms of the fibers are increased by both the characteristic morphology with an elongated rather than a circular section[16] and a complex spongy polymeric structure, leading to filaments with a porosity that is higher than that considered in the past, as evidenced from the SEM images. To investigate the RL emission, the dyed fibers are deposited on quartz and pumped.

The aim of our work is the observation of how the stimulated emission is distributed along the network and in which way it can be controlled by changing appropriate experimental parameters, for instance, the pumping energy.

A sketch of the experimental setup is illustrated in Figure 1b. A Q-switched Nd:Yag laser operating at 532 nm, 4 ns pulse duration, and 10 Hz repetition rate is used as the pump; a motorized rotating $\lambda/2$ plate with a polarized cube beam splitter are used for automatic energy variation. The laser light is then 50% splitted and the reflected portion is detected by an energy meter to monitor the pump energy. The transmitted beam is focalized with an objective OBJ1 to a diameter waist of the Gaussian profile of 180 μm on the sample. The emission from the pumped area is magnified with 50× magnification by an objective OBJ2 with numerical aperture 0.4 and divided into two parts by a beamsplitter BS for both fluorescence imaging on a charged camera device (CCD) and spectral detection. A filter is used for removing residual pump light. A fiber is mounted on motorized stages to scan the whole emission and connected to a Horiba Jobin Yvon spectrometer equipped with an electrically cooled Symphony CCD array detector. The same emission is imaged on a Retiga R1 CCD camera. Fiber and CCD are placed at the same distance $L$=21 cm in order to have the same magnification (see sketch in Figure 1b). Both CCD and spectrometer are triggered by the flash-lamp of the pump laser. A calibration tool is used to have the exact correspondence between each position of the fiber and each point of the CCD display with the same spatial resolution of 0.1 μm in the sample plane.

An example of the fluorescence image is reported in Figure 1c, and in Figure 1d representative spectra taken at different spatial coordinates inside the fibers show a broad band typical of amplified spontaneous emission (ASE) and RL peaks on the top. The coexistence of ASE and RL emission has been observed in other planar and wave guiding systems.[13,17,18] The growth of the intensity by increasing input energy is shown in Figure 1e, and the random laser energy threshold in the range 0.3-1 μJ is estimated from different points of the network, both from the recorded intensity on the camera and the correspondent spectral amplitude. These results are reported in the Supporting Information, Figure S1 and also clearly show the accordance between the trends of the emitted intensity from different points of the considered network recorded simultaneously by the fiber connected to the spectrograph and the CCD. We performed such validation tests on different samples and on different points. This verification is necessary because, to quantify the spatial distribution of the RL emission and its activity in this work, we use fluorescence imaging that allows us to have access to the emitted intensity of various regions of the entire pumped area simultaneously at any single shot. Instead, the point by point spectral investigation is based on moving the fiber and recording data at many different positions, each one corresponding to a different pumping pulse. This is also legitimated by the fact that our study is based on the analysis of the emission intensities, and the spectral shape does not affect the results.

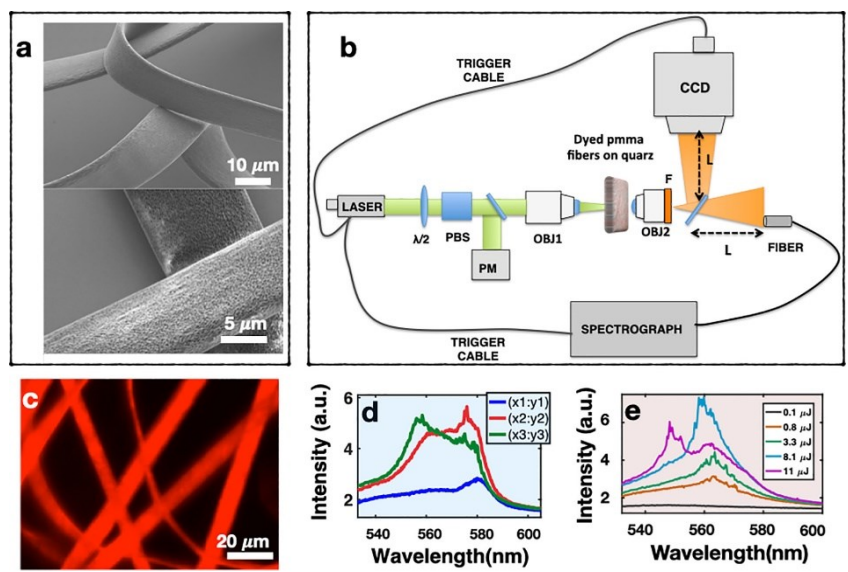

**Figure 1**. Sample images, experimental setup, and RL emission spectra. (a) SEM images of the complex porous network. (b) Sketch of the RL setup: a pump pulsed laser at 532 nm is focalized by objective OBJ1 on the PMMA dyed fibers deposited on quartz and the fluorescence is collected by objective OBJ2 and splitted in two. Half is recorded by a CCD and half is scanned by a moving fiber connected to spectrograph. The distance $L$ of the detectors from OBJ2 is fixed, giving a 50× magnification. For input energy tuning a $\lambda/2$ coupled with polarized cubed beam splitter (PBS) is used. A 50% beam splitter is used to record the exact energy by a power meter PM. A colored filter F is mounted on OBJ2 to remove residual pumping light. (c) Fluorescence image in false colors of a representative dyed polymeric network. (d) Different spectra taken at various points with coordinate ($x_k$, $y_k$) inside the network arms showing RL emission. (e) Representative emission spectra at diverse pumping energy showing a random laser.

Indeed, strong shot to shot intensity fluctuations measured at high pumping with interesting variability from different regions of the illuminated network are visible in the Movie S2 and plotted in Figure S2 of the Supporting Information. It is evident that at the same energy and under identical experimental conditions some regions have emission fluctuations much stronger than that of other ones.

We investigate this intriguing behavior by recording 100 single shot images at different input energies, and we estimate for each pixel the mean intensity and the variance. The correspondent obtained maps at various pumping are shown in Figure 2a.

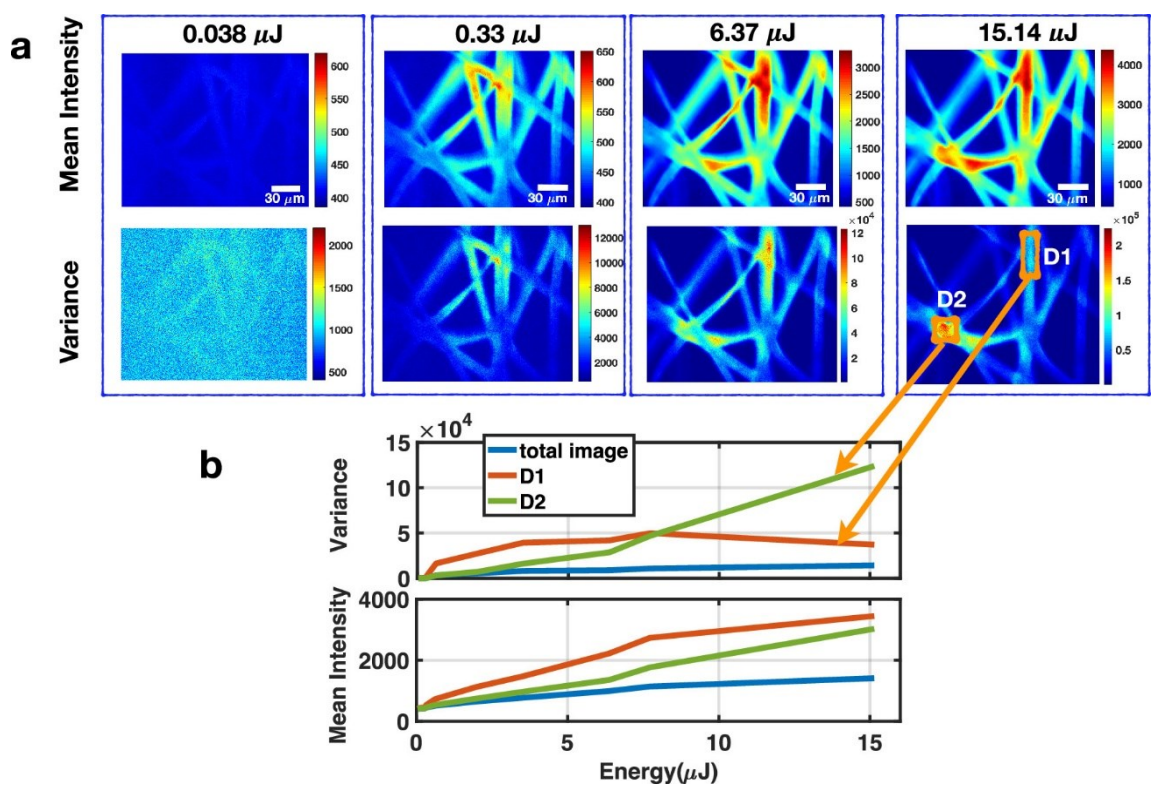

**Figure 2**. Maps of emitted mean intensity and variance. (a) Mean intensity and variance calculated from 100 single shot frames and mapped for various input laser energies. The calibration bars report the values as calculated from those exactly recorded on the CCD for each pixel without autoscales and normalizations. (b) Energy trends of the mean intensity and variance averaged over entire maps of panel (a) (blue line) and over two distinct domains, D1 and D2, indicated in panel a with orange squares. Error bars are inside the lines.

At low energy (left panel in Figure 2a), the spontaneous emission is homogeneously distributed along the fibers and does not present spatial heterogeneity in the fluctuations, as shown from the variance map. By increasing pumping, the average emission starts to be different in diverse points, and the regions with higher emission intensity also present high intensity fluctuations, this can be seen in Figure 2a at 0.33 µJ. For higher injected energy, a different behavior is observed: some domains of the active network have high mean emission intensity and high variance, while others, despite the strong brightness, are less fluctuating and more stable in time. We show this more clearly in Figure 2b, where variance and mean intensity from two selected domains, D1 and D2, are plotted. The two regions present a growing mean emission by increasing the input energy, while the shot to shot fluctuations are divergent after 7.5 µJ. Both D1 and D2 are intense, but D2 is much more blinking at certain input energies. We observe this unexpected behavior in other parts of the network and also in other similar samples, and we see that it is typical of the complex mesh of ribbon-like porous fibers. In Figure 2b we also plot mean intensity and variance averaged over all pixels of the images, and both present a growing trend for increasing pump. This shows that the emission from single parts of the network does not follow that from the whole system.

In addition, we exclude the degradation of micrometric regions by monitoring their mean emitted intensity. A clear signature of deterioration is the irreversible and not fluctuating dropping down of the mean emitted intensity from single points, that may occur at high pumping. An example is shown in Figure S1 of the Supporting Information. The results reported in this work are well below such a damage point.

**Replica Symmetry Breaking Maps**

To better understand the nature of the bright regions and for a visualization of their activity, we built replica symmetry breaking (RSB) maps from the fluorescence images. Replica symmetry breaking in random lasers has been introduced in 2015[17] and widely demonstrated in the last years both theoretically and experimentally for several kinds of RLs.[19-26] It has been identified at and above the laser threshold and quantified by the bimodal distribution of the Parisi overlap parameter.[27,28]

For our investigation, we calculate the overlap parameter directly from the images. We record 100 single shot frames at fixed energy, and each one is a replica. We divide each frame of 1024×1376 pixels into squares of 16×16 that we name macropixel, and we calculate the overlap $q_{\alpha,\beta}^j$ for each macropixel as

$$q_{\alpha,\beta}^j = \frac{\sum_{x_k}\sum_{y_k} \Delta_{\alpha,j}(x_k,y_k)\Delta_{\beta,j}(x_k,y_k)}{\sqrt{\sum_{x_k}\sum_{y_k} \Delta_{\alpha,j}(x_k,y_k)^2}\sqrt{\sum_{x_k}\sum_{y_k} \Delta_{\beta,j}(x_k,y_k)^2}} \quad (1)$$

where $\alpha$ and $\beta$ are replica indexes, $j$ is the macropixel index, and $x_k$ and $y_k$ are pixel indexes inside one macropixel; in our case, they run from 1 to 16. In Eq. (1)

$$\Delta_{\alpha,j}(x_k,y_k) = I_\alpha^j(x_k,y_k) - \bar{I}^j(x_k,y_k) \quad (2)$$

with $I_\alpha^j(x_k, y_k)$ being the intensity at the pixel with coordinate $(x_k, y_k)$ of $j$th macropixel for replica $\alpha$ and

$$\bar{I}^j(x_k,y_k) = \frac{1}{N_{shot}} \sum_{\alpha=1}^{N_{shot}} I_\alpha^j(x_k,y_k) \quad (3)$$

as the intensity at the pixel with coordinate $(x_k, y_k)$ of $j$th macropixel averaged over all 100 replica (shots).

We choose macropixels of 16×16 pixels because this results in a good range for the sums in Eq. (1) and the estimation of $q_{\alpha,\beta}^j$ and it allows us to have a good resolution in the RSB maps reported in the following.

By this procedure from the images we have access to the sets of all $N_s(N_s-1)/2$ values $q^j$ of $q_{\alpha,\beta}^j$ and we determine the distributions $P(q^j)$ for all 64×86 macropixels $j$ and for each pumping energy. Representative $P(q)$ distributions at different energies and for two macropixels inside the domains D1 and D2 of Figure 2 are depicted in Figure 3a. At low energy, the distributions are centered around $q=0$, meaning that both points are in the paramagnetic regime of the spontaneous emission. By increasing pumping, the distributions broaden, and after a certain threshold, bimodal shapes around the values $q=\pm 1$ are obtained, typical of RL out of equilibrium regime.[17] At high pumping, a different and novel scenario emerges. In the macropixel inside region D1, the distribution $P(q)$ loses the bimodality, while the one in D2 does not present significative change.

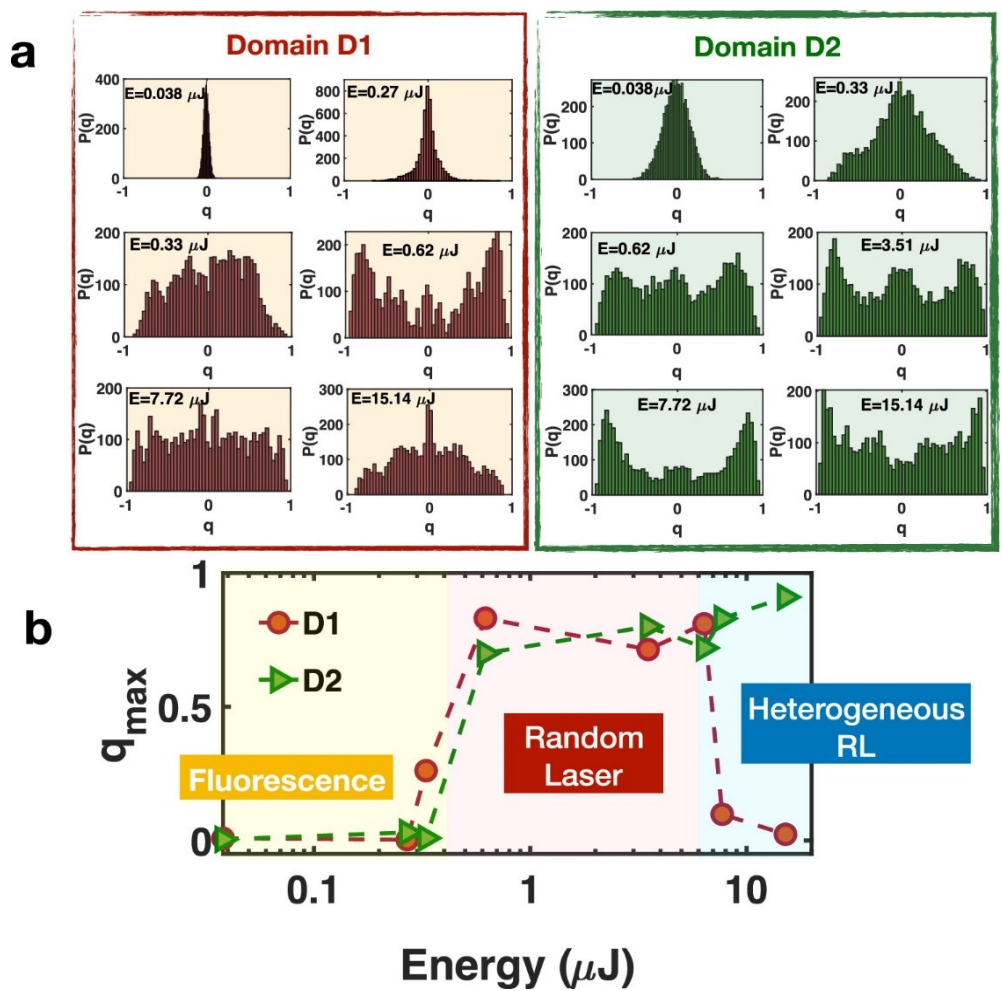

**Figure 3**. Overlap distributions evidencing heterogeneous RL. (a) Distributions of the overlap parameter $q$ at various input energies and for two distinct domains D1 and D2, indicated in Figure 2. (b) $q_{max}$ versus pump energy for D1 and D2 regions, showing three diverse regimes of the emission.

Following the previous works, we choose $q_{max}$, which is the absolute value of $q$ for which $P(q)$ is maximum, as a suitable parameter to define the different regimes and transitions. In Figure 3b we report plots of $q_{max}$ versus input energy for the macropixels inside domains D1 and D2 and estimated from the correspondent distributions of Figure 3a. The transition from spontaneous to stimulated emission at the RL threshold energy $E_{th} \simeq 0.4$ µJ is defined by the value $q_{max}$=0.5. This estimation is in good agreement with the threshold measured from spectral emissions (Figure S1 in the Supporting Information). At higher energy ($E \geq 7$ µJ), an unexpected divergence in $q_{max}$ profile is clearly shown. The point with $q_{max} \simeq 1$ continues to lase, while the other with $q_{max} \simeq 0$ switches off. This determines for the first time the appearance of spatial heterogeneity in the RL emission at high pumping and is the novelty of our research.

We stress that this dumping of the RL activity is not due to any sample damage or experimental variation, as shown in the mean intensity maps in Figure 2. At the input energy of 15.14 µJ both domains D1 and D2 are bright, only the entity of the shot to shot intensity fluctuations, directly connected to the RL action, is diverse. The different regions of the network are equally shiny because of the presence of ASE, but the stimulated emission has a switching activity due to mode competition between the interconnected domains.

We may also notice that the narrowing of $P(q)$ well above the RL threshold has been previously reported,[24] but the authors in this work do not give further and deeper explanation of the observed suppression of the RL glassy regimes and refer to future studies.

We calculate $q_{max}$ for all macropixels of the images at various pumping, and we report in Figure 4 the obtained RSB maps. At low energy, $q_{max} \leq 0.06$ and varies around the zero value in all points, and the whole system is in the fluorescence paramagnetic regime. At 0.33 µJ, a nucleation center of RL emission, defined by $q_{max} \simeq 0.6$, takes place. Above this energy other different parts of the heterogeneous RL are switched on and their activity changes randomly at different pumps. On the maps we indicate some of the ON ($q_{max}$>0.5) and OFF ($q_{max}$<0.5) regions, and their dynamics is clearly evidenced.

It is noticeable that some regions alternate ON and OFF states as the one in the lower part of the maps. This specific region is in state ON at 0.62 and 3.51 µJ, in state OFF at 6.37 and 7.72 µJ, and again in state ON at 15.14 µJ. Such behavior depends on the specific RL region and on the injected energy. Moreover, same kind of analysis on single shot measurements with different temporal delays in the range 0.5-20 s reproduces the heterogeneous spatial distribution of RL domains and their switching activity observed with 10 Hz repetition rate, confirming that the observed blinking effect is not due to any memory effect between subsequent pumping pulses.

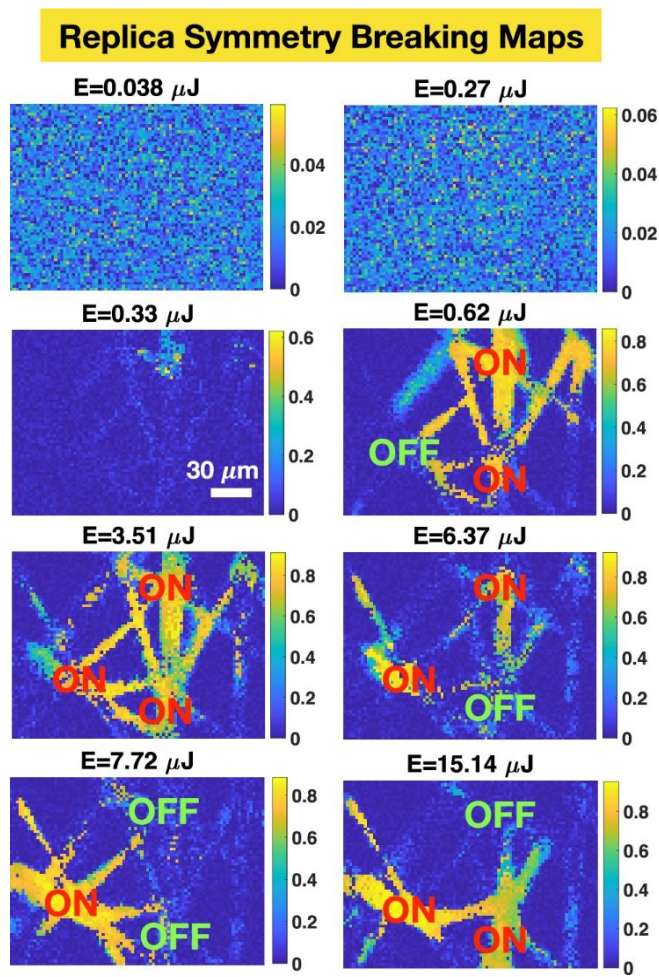

**Figure 4**. Replica symmetry breaking maps. Maps of the $q_{max}$ values calculated for each macropixel at different input energies. ON and OFF indicate the activation status of the RL emission.

The maps in Figure 4 show unambiguously which parts of the network have broken replica symmetry and are in the glassy-RL state and which parts have only spontaneous emission at distinct pumpings. The mapping of $q_{max}$, that is, a parameter that can assume values between 0 and 1 and defines the transition to the laser regime, allows the visualization of heterogeneous RL with switching and variable activity with 2 μm spatial resolution given by the length of a macropixel.

We may stress that the observed RL switching of the domains is characteristic of networks made of ribbon-like and porous PMMA fibers. Indeed, we have investigated the emission from webs of PMMA fibers of different morphologies, structures, and thicknesses. In networks with fibers with a circular section of 10-20 μm in diameter and less porosity and with fibers of a thinner circular section (1-2 μm in diameter) and less porosity, we observe random laser emission mainly from the nodes without any evidence of switching. Above threshold all nodes remain in the nonergodic state with $q_{max}\simeq1$. We think that the reported blinking effect is due to the high porosity and the planar shape of the fibers responsible or RL emission along the arms of the network. The emitted regions are wider and closer compared to systems where the radiation is confined mainly at the nodes.

Before concluding this section, we point out that not in all RLs the condition $q_{max}\simeq0$ above laser threshold indicates the ASE state, that here we name OFF state. There are examples of RLs that do not show relevant intensity fluctuations and replica symmetry breaking at any high input energy.[17] In these systems ASE in not present and the emission is sharp, stable, spatially confined, and without fluctuations at the same energy and stability as others, and switching between these states. For these reasons we exclude that in our networks the condition $q_{max}\simeq0$ at high pumping indicates stable RL. We also exclude fluorescence because we did not observe a signal decrease in these domains and only at certain input energies. Contrarily, ASE, that is typical of planar systems, shows a $q_{max}\simeq0$, and it is characterized as a smooth, though strong, stable emission.

In the investigated networks, the activation of RL changes from one region to another because the regions are connected and in competition thanks to wave guiding. This provokes the switching between ASE (OFF) and RL (ON) states. Further investigations by using streak cameras with picosecond/femtosecond temporal resolution would help to better understand the described complex behavior and the true role of mode interference by analyzing optical fluctuations in the time domain.

**Pearson Correlation Curves**

The next step of our investigation is the evaluation of the spatial extension of the RL domains and of the correlation among them. To this aim we calculate the Pearson correlation coefficient[13,29] between all macropixels as

$$C_{i,j} = \frac{\sum_{\alpha=1}^{N_{shot}}[I_\alpha^i - \bar{I}^i][I_\alpha^j - \bar{I}^j]}{\sqrt{\sum_{\alpha=1}^{N_{shot}}[I_\alpha^i - \bar{I}^i]^2}\sqrt{\sum_{\alpha=1}^{N_{shot}}[I_\alpha^j - \bar{I}^j]^2}} \quad (4)$$

where $i$ and $j$ are indexes of macropixel and $\bar{I}^i$ represents the average intensity over 100 shots for macropixel $i$. If two macropixels have $C_{i,j}$ close to zero, their fluctuations are uncorrelated and thus do not belong to the same RL domain. While correlated macropixels with $C_{i,j} \simeq 1$ can be considered from the same lasing region because they are similar in the intensity fluctuations, this is an indication of RL occurrence.

In Figure 5 we show the correlation curves at different input energies and along four representative lines depicted in Figure 5a.

The dependence on the pumping is shown in Figure 5b for one of the selected directions. At low energy in the fluorescence and initial RL regime, the points along the arm are highly correlated, while by growing pumping a distinct decrease is evidenced for distances above 40 μm. This allows us to determine the spatial extension of the RL domains shown in the RSB maps. We estimate an average radial size of (50±10) μm. Moreover, the Pearson correlation trends highlight a more complex scenario. This is illustrated in Figure 5c where the four different lines show two distinct behaviors: the decays of lines 1 and 2 signal the existence of confined RL domains, while lines 3 and 4 have only slightly decreased correlated emission, showing activated RL along entire linkers.

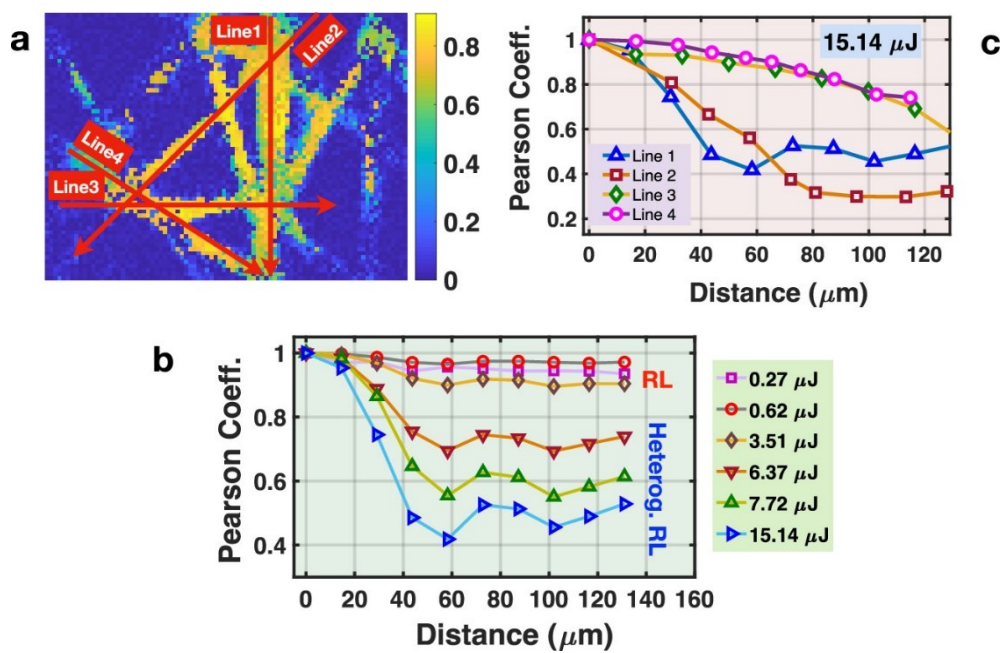

**Figure 5**. Pearson correlation curves. (a) Four selected lines along which the correlation coefficient is estimated. The Pearson correlation is calculated between the first point of each arrow and the other points until the end on the line. The image is the RSB map at the energy 3.51 μJ of the homogeneous RL regime. (b) Correlation curves of line 1 at various pumping showing the emergence of correlated domains at high energy. (c) Correlation curves of the four selected lines at the highest energy evidencing the coexistence of isolated RL domains and more elongated ones in the heterogeneous disordered laser.

**Conclusion**

In conclusion, we fabricate a complex emitter network with switching micrometric domains activated by the pumping intensity. We visualize such an effect for the first time by building RSB maps, a robust tool that can be used in other types of RLs by simply imaging the near-field fluorescence and quantifying the single pixel shot to shot intensity fluctuations. Our finding is an early demonstration of miniaturized laser network with heterogeneous patterns. The emission of such devices can be tuned spatially by input beam shaping and ad-hoc composition and topology of wave guides for applications in sensing, spectroscopy and structured illumination microscopy.[30-32]

ASSOCIATED CONTENT

**Supporting Information**

Demonstration of the accordance between spectral and imaging detection. Evidence of spatial heterogeneity in the intensity fluctuations

Movie S2: single shot frames showing the variability of the emission intensity fluctuations in the different points of the network at the high energy of 15.14 μJ

ACKNOWLEDGMENT

The author thanks MD Deen Islam for technical support in the photonics laboratory. Luigi Romano is gratefully acknowledged for SEM images. The research leading to these results has received funding from CNR Bilateral Joint Italy-Russia Project 2018-2021; ELITE Project Prot. 85-2017-15111-Avviso Pubblico "Progetti di Gruppi di Ricerca - Conoscenza e Cooperazione per un nuovo modello di sviluppo" Legge 13/2008-264 art. 4 (Codice CUP B86C17000460002) - Regione Lazio - Lazio Innova; Italian Minister of University and Research PRIN Project 201795SBA3.

REFERENCES

(1) Andrady, A. L. *Science and Technology of Polymer Nanofibers*; A. John Wiley & Sons, Inc. Hoboken, USA, 2008.

(2) O'Carroll, D.; Iacopino, D.; O'Riordan, A.; Lovera, P.; O'Connor, E.; O'Brien, G.; Redmond, G. Poly(9,9-dioctylfluorene) Nanowires with Pronounced *β*-Phase Morphology: Synthesis, Characterization, and Optical Properties. *Adv. Mater.* **2008**, *20*, 42–48.

Published in ACS Photonics, doi: 10.1021/acsphotonics.0c01803 (2021).

(3) Wang, P.; Wang, Y.; Tong, L. Functionalized Polymer Nanofibers: a Versatile Platform for Manipulating Light at the Nanoscale. *Light Sci. Appl*. **2013**, *2*, e102.

(4) Tu, D.; Pagliara, S.; Cingolani, R.; Pisignano, D. An Electrospun Fiber Phototransistor by the Conjugated Polymer Poly[2-methoxy-5-(2'-ethylhexyloxy)-1,4-phenylenevinylene]. *Appl. Phys. Lett*. **2011**, *98*, 023307.

(5) Gaio, M.; Moffa, M.; Castro-Lopez, M.; Pisignano, D.; Camposeo, A.; Sapienza, R. Modal Coupling of Single Photon Emitters Within Nanofiber Waveguides. *ACS Nano* **2016**, *10*, 6125–6130.

(6) Gaio, M.; Saxena, D.; Bertolotti, J.; Pisignano, D.; Camposeo, A.; Sapienza, R. A Nanophotonic Laser on a Graph. *Nat. Comm.* **2019**, *10*, 226.

(7) Cao, H. Lasing in Random Media. *Waves in Random Media and Complex Media* **2003**, *13*, R1–R39.

(8) Wiersma, D. S. The Physics and Applications of Random Lasers. *Nat. Phys.* **2008**, *4*, 359.

(9) Sznitko, L.; Romano, L.; Wawrzynczyk, D.; Cyprych, K.; Mysliwiec, J.; Pisignano, D. Stacked Electrospun Polymer Nanofiber Heterostructures with Tailored Stimulated Emission. *RSC Adv.* **2018**, *8*, 24175–24181.

(10) Lepri, S.; Trono, C.; Giacomelli, G. Complex Active Optical Networks as a New Laser Concept. *Phys. Rev. Lett.* **2017**, *118*, 123901.

(11) Giacomelli, G.; Lepri, S.; Trono, C. Optical Networks as Complex Lasers. *Phys. Rev. A* **2019**, *99*, 023841.

(12) Resta, V.; Camposeo, A.; Montinaro, M.; Moffa, M.; Kazlauskas, K.; Jursenas, S.; Tomkeviciene, A.; Grazulevicius, J. V.; Pisignano, D. Nanoparticle-Doped Electrospun Fiber Random Lasers with Spatially Extended Light Modes. *Opt. Express* **2017**, *25*, 24604–24614.

(13) Montinaro, M.; Resta, V.; Camposeo, A.; Moffa, M.; Morello, G.; Persano, L.; Kazlauskas, K.; Jursenas, S.; Tomkeviciene, A.; Grazulevicius, J. V.; Pisignano, D. Diverse Regimes of Mode Intensity Correlation in Nanofiber Random Lasers through Nanoparticle Doping. *ACS Photonics* **2018**, *5*, 1026–1033.

(14) Zhang, R.; Knitter, S.; Liew, S. F.; Omenetto, F. G.; Reinhard, B. M.; Cao, H.; Dal Negro, L. Plasmon-Enhanced Random Lasing in Bio-Compatible Networks of Cellulose Nanofibers. *Appl. Phys. Lett.* **2016**, *108*, 011103.

(15) Xia, J.; He, J.; Xie, K.; Zhang, X.; Hu, L.; Li, Y.; Chen, X.; Ma, J.; Wen, J.; Chen, J.; Pan, Q.; Zhang, J.; Vatnik, I. D.; Churkin, D.; Hu, Z. Replica Symmetry Breaking in FRET-Assisted Random Laser Based on Electrospun Polymer Fiber. *Annalen der Physik* **2019**, *531*, 1900066.

(16) Ghofraniha, N.; Viola, I.; Zacheo, A.; Arima, V.; Gigli, G.; Conti, C. Transition from Nonresonant to Resonant Random Lasers by the Geometrical Confinement of Disorder. *Opt. Lett.* **2013**, *38*, 5043–5046.

(17) Ghofraniha, N.; Viola, I.; Dimaria, F.; Barbarella, G.; Gigli, G.; Leuzzi, L.; Conti, C. Experimental Evidence of Replica Symmetry Breaking in Random Lasers. *Nat. Comm.* **2015**, *6*, 6058.

(18) Anni, M.; Lattante, S.; Stomeo, T.; Cingolani, R.; Gigli, G.; Barbarella, G.; Favaretto, L. Modes Interaction and Light Transport in Bidimensional Organic Random Lasers in the Weak Scattering Limit. *Phys. Rev. B* **2004**, *70*, 195216.

(19) Antenucci, F.; Crisanti, A.; Leuzzi, L. The Glassy Random Laser: Replica Symmetry Breaking in the Intensity Fluctuations of Emission Spectra. *Sci. Rep.* **2015**, *5*, 16792.

(20) Antenucci, F.; Conti, C.; Crisanti, A.; Leuzzi, L. General Phase Diagram of Multimodal Ordered and Disordered Lasers in Closed and Open Cavities. *Phys. Rev. Lett.* **2015**, *114*, 043901.

(21) Gomes, A. S. L.; Lima, B. C.; Pincheira, P. I. R.; Moura, A. L.; Gagné, M.; Raposo, E. P.; de Araújo, C. B.; Kashyap, R. Glassy Behavior in a One-Dimensional Continuous-Wave Erbium-Doped Random Fiber Laser. *Phys. Rev. A* **2016**, *94*, 011801.

(22) Tommasi, F.; Ignesti, E.; Lepri, S.; Cavalieri, S. Robustness of Replica Symmetry Breaking Phenomenology in Random Laser. *Sci. Rep.* **2016**, *6*, 37113.

(23) Pincheira, P. I. R.; Silva, A. F.; Fewo, S. I.; no, S. J. M. C.; Moura, A. L.; Raposo, E. P.; Gomes, A. S. L.; de Araújo, C. B. Observation of Photonic Paramagnetic to Spin-Glass Transition in a Specially Designed $TiO_2$ Particle-Based Dye-Colloidal Random Laser. *Opt. Lett.* **2016**, *41*, 3459–3462.

(24) Gomes, A. S. L.; Raposo, E. P.; Moura, A. L.; Fewo, S. I.; Pincheira, P. I. R.; Jerez, V.; Maia, L. J. Q.; de Araújo, C. B. Observation of Lévy Distribution and Replica Symmetry Breaking in Random Lasers from a Single Set of Measurements. *Sci. Rep.* **2016**, *6*, 27987.

(25) Basak, S.; Blanco, A.; López, C. Large Fluctuations at the Lasing Threshold of Dolid- and Liquid-State Dye Lasers. *Sci. Rep.* **2016**, *6*, 32134.

(26) Sarkar, A.; Bhaktha, B. N. S.; Andreasen, J. Replica Symmetry Breaking in a Weakly Scattering Optofluidic Random Laser. *Sci. Rep.* **2020**, *10*, 2628.

(27) Parisi, G. Order Parameter for Spin-Glasses. *Phys. Rev. Lett.* **1983**, *50*, 1946–1948.

(28) Parisi, G. The Order Parameter for Spin Glasses: a Function on the Interval 0-1. *J. Phys. A* **1980**, *13*, 1101.

(29) Leonetti, M.; Conti, C.; López, C. Dynamics of Phase-Locking Random Lasers. *Phys. Rev. A* **2013**, *88*, 043834.

(30) Boschetti, A.; Taschin, A.; Bartolini, P.; Tiwari, A. K.; Pattelli, L.; Torre, R.; Wiersma, D. S. Spectral Super-Resolution Spectroscopy Using a Random Laser. *Nat. Photon.* **2020**, *14*, 177.

(31) Redding, B.; Choma, M. A.; Cao, H. Speckle-Free Laser Imaging Using Random Laser Illumination. *Nat. Phot.* **2012**, *6*, 355.

(32) Lee, M.; Callard, S.; Seassal, C.; Jeon, H. Taming of Random Lasers. *Nat. Photon.* **2020**, *13*, 445.

# Supporting Information

# Heterogeneous Random Laser with Switching Activity Visualized by Replica Symmetry Breaking Maps

Loredana M. Massaro,[†] Silvia Gentilini,[‡] Alberto Portone,[¶] Andrea Camposeo,[¶] Dario Pisignano,[§] Claudio Conti,[‡] Neda Ghofraniha[*,‡]

[†]*Dipartimento di Fisica, Università La Sapienza, P.le A. Moro 5, I-00185, Rome, Italy*

[‡]*Istituto dei Sistemi Complessi-CNR, UOS Università La Sapienza, P. le A. Moro 2, I-00185, Rome, Italy*

[¶]*NEST, Istituto Nanoscienze-CNR and Scuola Normale Superiore, Piazza S. Silvestro 12, I-56127 Pisa, Italy*

[§]*NEST, Istituto Nanoscienze-CNR and Dipartimento di Fisica, Università di Pisa, I-56127 Pisa, Italy*

[*]E-mail: neda.ghofraniha@cnr.it

**Accordance between spectral and imaging detection**

In figure S1 we report plots of the normalised emission intensity versus pumping energy from two points P1 and P2 inside the network, as shown in the related image, measured by the spectrograph and the CCD camera. Each value is an average over 100 acquisitions. The accordance is evident and the slight discrepancy is due to the different accumulation numbers used in the two techniques: images are single shot measurements and the spectra are averaged over 10 shots, being each spectrum acquired with an exposure time of 1 second. In the inset same plot in log-scale zooms on the low

Published in ACS Photonics, doi: [10.1021/acsphotonics.0c01803](10.1021/acsphotonics.0c01803) (2021).energy trends showing the threshold between spontaneous and stimulated emission. Additionally, the mean intensity trends are used in the calibration procedure to exclude sample deterioration at high energy. Indeed, in the reported example the decrease of the emission at high energy (> 7 µJ) in P1 and P2 shows the irreversible dropping down of emission inside the network and is indicative of dye bleaching at high pumping. By this method we monitor point by point sample degradation in all the investigations. The analysis reported in Figure S1 is an example of calibrations we performed. Similar trends were obtained on different samples and different points, validating the experimental procedures reported in the manuscript.

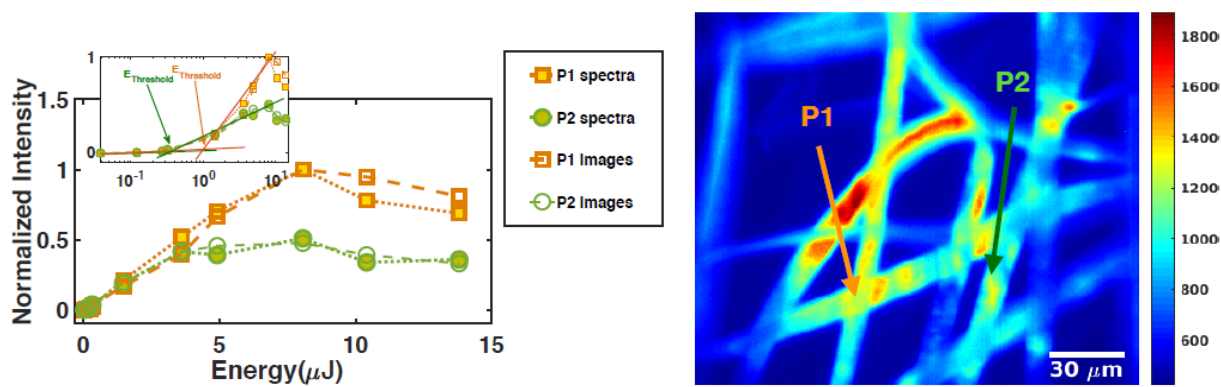

**FIGURE S1**: Comparison between emitted intensity taken from spectra and images on the CCD. Left: Normalised emitted intensity versus input energy from two distinct points P1 and P2 in the network, measured by the spectrograph and the CCD camera. The intensity on the spectrometer is the normalised spectral area and that from the images is the normalised intensity recorded by the pixels correspondent to P1 and P2. Inset: same plot in log-scale zooming on the low energy part and showing clearly the RL threshold of both P1 and P2. Error bars are inside the symbols. Right: The image of the selected network used for validation tests, P1 and P2 are marked.

**Evidence of spatial heterogeneity in the intensity fluctuations**

The movie S2 with single shot frames shows the variability of the emission intensity fluctuations in the different points of the network at the high energy of 15.14 µJ. Regions D1 and D2 are indicated and strong fluctuations in separated and distinct regions are signature of heterogeneous random laser. For the sake of major clarity we report three images at single shots taken from movie S2 and plots of the emission intensities in the domains D1 and D2 at different shots. It is evident that D2 has more variable pulse to pulse emissions, as also demonstrated by the calculated variances reported in Figure 2 of the manuscript.

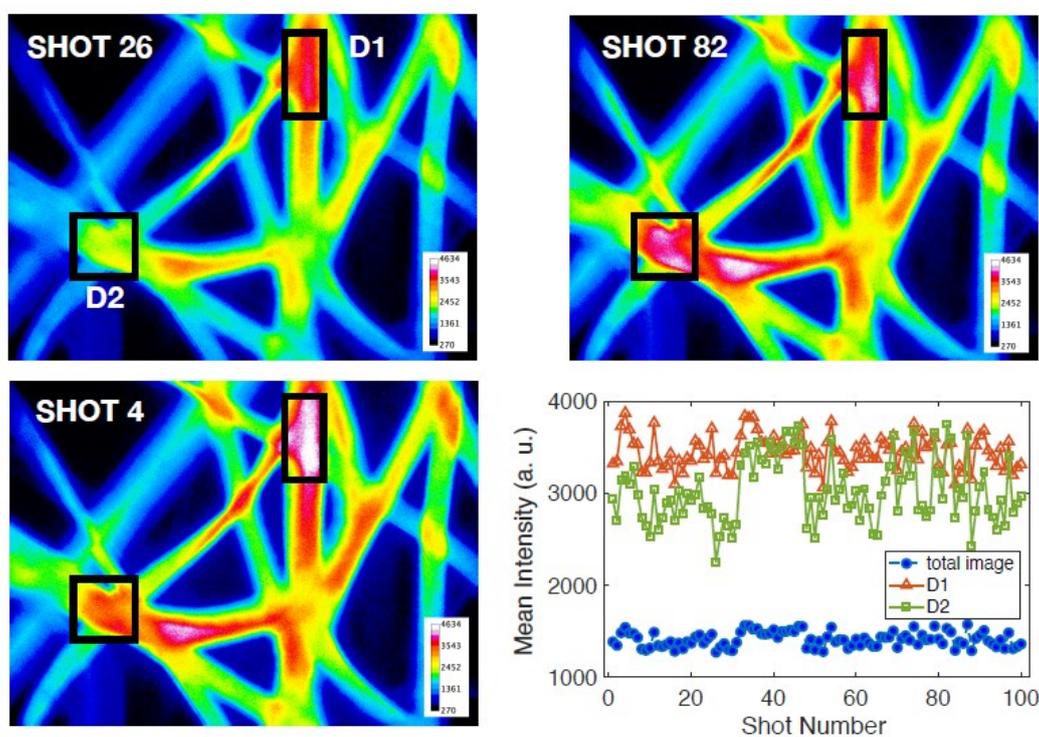

**FIGURE S2**: Evidence of different intensity fluctuations. Three images at single shots taken from movie S2 and plots of the emission intensity calculated from the movie S2 in regions D1, D2 and from whole the network. Error bars are inside the symbols.

**Movie S2**: single shot frames showing the variability of the emission intensity fluctuations in the different points of the network at the high energy of 15.14 µJ.